\documentclass[aps,prl,twocolumn,amsmath,amssymb,nofootinbib,superscriptaddress]{revtex4-2}

\usepackage{graphicx}
\usepackage[colorlinks]{hyperref}
\usepackage{textcomp}
\usepackage{color,soul}
\usepackage{setspace}

\begin{document}

\title{Identification of Light Sources using Machine Learning}

\author{Chenglong You}
\email[]{cyou2@lsu.edu}
\affiliation{Quantum Photonics Laboratory, Department of Physics and Astronomy, Louisiana State University, Baton Rouge, LA 70803, USA}

 \author{Mario A. Quiroz-Ju\'arez}
  \affiliation{Instituto de Ciencias Nucleares, Universidad Nacional Aut\'onoma de M\'exico, Apartado Postal 70-543, 04510 Cd. Mx., M\'exico}
  
 \author{Aidan Lambert}
 \affiliation{Quantum Photonics Laboratory, Department of Physics and Astronomy, Louisiana State University, Baton Rouge, LA 70803, USA}

 \author{Narayan Bhusal}
 \affiliation{Quantum Photonics Laboratory, Department of Physics and Astronomy, Louisiana State University, Baton Rouge, LA 70803, USA}

 \author{Chao Dong}
 \affiliation{Quantum Photonics Laboratory, Department of Physics and Astronomy, Louisiana State University, Baton Rouge, LA 70803, USA}

 \author{Armando Perez-Leija}
  \affiliation{Max-Born-Institut, Max-Born-Stra{\ss}e 2A, 12489 Berlin, Germany}

 \author{Amir Javaid}
  \affiliation{Quantum Photonics Laboratory, Department of Physics and Astronomy, Louisiana State University, Baton Rouge, LA 70803, USA}

 \author{Roberto de J. Le\'on-Montiel}
  \affiliation{Instituto de Ciencias Nucleares, Universidad Nacional Aut\'onoma de M\'exico, Apartado Postal 70-543, 04510 Cd. Mx., M\'exico}

\author{Omar S. Maga\~{n}a-Loaiza}
 \affiliation{Quantum Photonics Laboratory, Department of Physics and Astronomy, Louisiana State University, Baton Rouge, LA 70803, USA}

%% To be edited by editor
% \dates{Compiled \today}
%\ociscodes{(140.3490) Lasers, distributed feedback; (060.2420) Fibers, polarization-maintaining;(060.3735) Fiber Bragg gratings.}
%% To be edited by editor
% \doi{\url{http://dx.doi.org/10.1364/XX.XX.XXXXXX}}

\begin{abstract}
The identification of light sources represents a task of utmost importance for the development of multiple photonic technologies. Over the last decades, the identification of light sources as diverse as sunlight, laser radiation and molecule fluorescence has relied on the collection of photon statistics or the implementation of quantum state tomography. In general, this task requires an extensive number of measurements to unveil the characteristic statistical fluctuations and correlation properties of light, particularly in the low-photon flux regime. In this article, we exploit the self-learning features of artificial neural networks and naive Bayes classifier to dramatically reduce the number of measurements required to discriminate thermal light from coherent light at the single-photon level. We demonstrate robust light identification with tens of measurements at mean photon numbers below one. Our work demonstrates an improvement in terms of the number of measurements of several orders of magnitude with respect to conventional schemes for characterization of light sources. Our work has important implications for multiple photonic technologies such as LIDAR and microscopy.
\end{abstract}

\maketitle

\section{Introduction}

The underlying statistical fluctuations of the electromagnetic field have been widely utilized to identify diverse sources of light \cite{glauber:63, mandel:95}.  In this regard, the Mandel parameter constitutes an important metric to characterize the excitation mode of the electromagnetic field and consequently to classify light sources \cite{mandel:79}. Similarly, the degree of optical coherence has also been extensively utilized to identify light sources \cite{mandel:79, mandel:65, liu:09, hlousek:19}. Despite the fundamental importance of these quantities, they require large amounts of data which impose practical limitations \cite{hlousek:19,dovrat:12, dovrat:13, zambra:05, howard:19}. This problem has been partially alleviated by incorporating statistical methods, such as bootstrapping, to predict unlikely events that are hard to measure experimentally \cite{dovrat:13, zambra:05, howard:19, hlousek:19}. Unfortunately, the constraints of these methods severely impact the realistic implementation of photonic technologies for metrology, imaging, remote sensing and microscopy \cite{howard:19,dowling:93, sher:18, wang:16, dowling:08, Omar:19}.
%Nevertheless, the successful classification of light through the Mandel parameter requires an accurate knowledge of photon statistics. Consequently, extremely large sets of data and multiple realizations of the experiment are required.

% sensitivity of the Mandel parameter to limited photon statistics produced by scarce data severely affects protocols in which the identification of photons emitted from different sources is crucial \cite{dovrat:13, zambra:05}.
% This constraint is

The potential of machine learning has motivated novel families of technologies that exploit self-learning and self-evolving features of artificial neural networks to solve a large variety of problems in different branches of science \cite{lecun:15, biamonte:17}. Conversely, quantum mechanical systems have provided new mechanisms to achieve quantum speedup in machine learning \cite{biamonte:17, Dunjko:16}. In the context of quantum optics, there has been an enormous interest in utilizing machine learning to optimize quantum resources in optical systems \cite{Hentschel:10, Lumino:18, Zeilinger:18}. As a tool to characterize quantum systems, machine learning has been successfully employed to reduce the number of measurements required to perform quantum state discrimination, quantum separability and quantum state tomography \cite{Lohani:18, Gao:18, Torlai:18}.

In this article, we demonstrate the potential of machine learning to perform discrimination of light sources at extremely low-light levels. This is achieved by training single artificial neurons with the statistical fluctuations that characterize coherent and thermal states of light. The self-learning features of artificial neurons enable the dramatic reduction in the number of measurements and the number of photons required to perform identification of light sources. For the first time, our experimental results demonstrate the possibility of using less than ten measurements to identify light sources with mean photon numbers below one. In addition, we demonstrate similar experimental results using the naive Bayes classifier, which are outperformed by our single neuron approach. Finally, we present a discussion on how a single artificial neuron based on an ADAptive LINear Element (ADALINE) model can dramatically reduce, by several orders of magnitude, the number of measurements required to discriminate signal photons from ambient photons. This possibility has strong implications for realistic implementation of LIDAR, remote sensing and microscopy.

\section{Experimental Setup and Model}

\begin{figure*}[!htb]
\centering\includegraphics[width=0.94\linewidth]{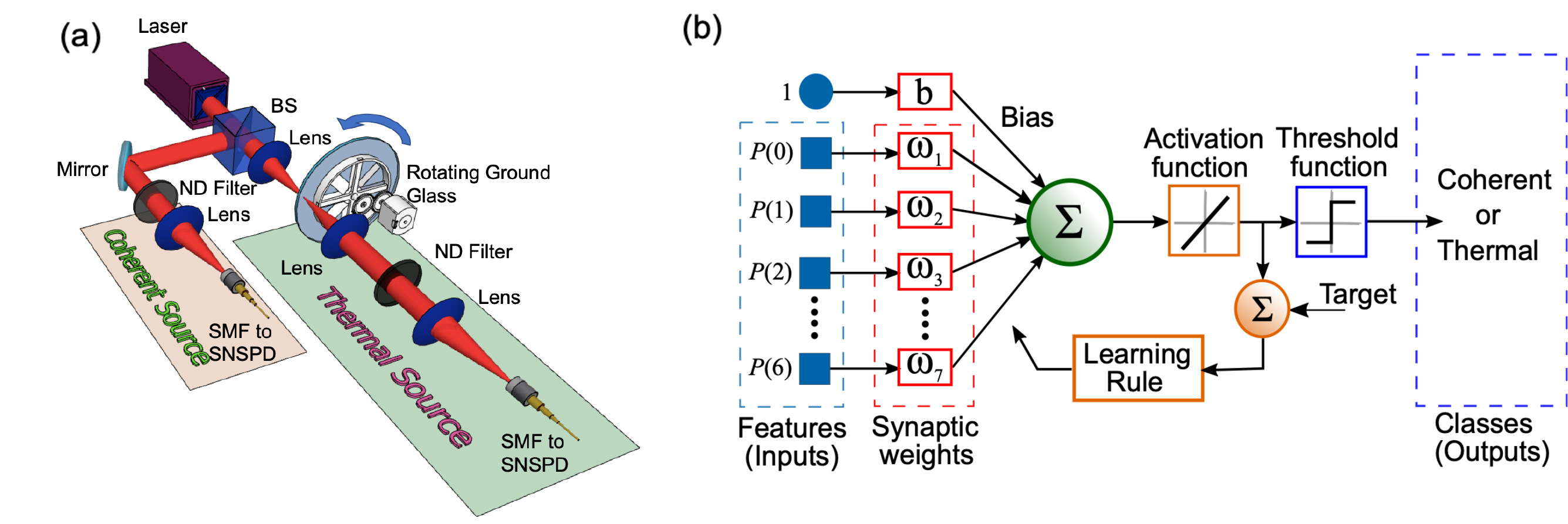}
\caption{(a) Schematic representation of the experimental setup. A laser beam is divided by a beam splitter (BS); the two replicas of the beam are used to generate light with Poissonian (coherent) and super-Poissonian (thermal) statistics. The thermal beam of light is generated by a rotating ground glass. Neutral density (ND) filters are utilized to attenuate light to the single-photon level. Coherent and thermal light beams are measured by superconducting nanowire single-photon detectors (SNSPDs). (b) Flow diagram of the ADALINE neuron used for demonstration of light source identification. Additional details are discussed in the body of the article.
 }
\label{fig:Exp_scheme}
\end{figure*}

As shown in Fig. \ref{fig:Exp_scheme} (a), we utilize a continuous-wave (CW) laser beam that is divided by a 50:50 beam splitter. The transmitted beam is focused onto a rotating ground glass which is used to generate pseudo-thermal light with super-Poissonian statistics. The beam emerging from the ground glass is collimated using a lens and attenuated by neutral-density (ND) filters to mean photon numbers below one. The attenuated beam is then coupled into a single-mode fiber (SMF). The fiber directs photons to a superconducting nanowire single-photon detector (SNSPD). The beam reflected by the beam splitter is used as a source of coherent light. This beam, characterized by Poissonian statistics, is also attenuated, coupled into a SMF and detected by a SNSPD. The brightness of the coherent beam is matched to that of the pseudo-thermal beam of light.

In order to perform photon counting from our SNSPDs, we use the surjective photon counting method described in ref. \cite{surjective}. The TTL pulses produced by our SNSPDs were detected and recorded by an oscilloscope. The data was divided in time bins of 1 \textmu s, which corresponds to the coherence time of our CW laser. Voltage peaks above \textasciitilde0.5 V were considered as one photon event. The number of photons (voltage peaks) in each time bin was counted to retrieve photon statistics. These events were then used for training and testing our ADALINE neuron and naive Bayes classifier.

The probability of finding $n$ photons in coherent light is given by $P_{\text{coh}}(n)= e^{-\bar{n}} (\bar{n}^n/n!)$, where $\bar{n}$ denotes the mean photon number of the beam. Furthermore, the photon statistics of thermal light is given by $P_{\text{th}}(n)=\bar{n}^n / (\bar{n}+1)^{n+1}$.
It is worth noting that the photon statistics of thermal light is characterized by random intensity fluctuations with a variance greater than the mean number of photons in the mode. For coherent light, the maximum photon-number probability sits around $\bar{n}$. For thermal light, the maximum is always at vacuum. However, when the mean photon number is low, the photon number distribution for both kinds of light becomes similar. Consequently, it becomes extremely difficult to discriminate one source from the other. The conventional approach to discriminate light sources makes use of histograms generated through the collection of millions of measurements \cite{dovrat:12,zambra:05,Burenkov:17,Montaut:18}.  Unfortunately, this method is not only time consuming, but also imposes practical limitations.

In order to dramatically reduce the number of measurements required to identify light sources, we make use of an ADALINE neuron. ADALINE is a single neural network model based on a linear processing element, proposed by Bernard Widrow \cite{Bernard}, for binary classification. In general, the neural networks undergo two stages: training and test. In the training stage, ADALINE is capable of learning the correct outputs (named as output labels or classes) from a set of inputs, so-called features, by using a supervised learning algorithm. In the test stage, this neuron produces the outputs of a set of inputs that were not in the training data, taking as reference the acquired experience in the training stage. Although we tested architectures far more complex than a single neuron for the identification of light sources, we found that a simple ADALINE offers a perfect balance between accuracy and simplicity (for more details, see the Supplementary Information). The structure of the ADALINE model is shown in Fig. \ref{fig:Exp_scheme}(b). The neuron input features are denoted by $P(n)$, which corresponds to the probability of detecting $n$ photons, in a single measurement event, for a given light source, namely coherent or thermal. Furthermore, the parameters $\omega_i$ are the synaptic weights and $b$ is a bias term.  In the training period, these parameters are optimized through the learning rule by using the error between the target output and neuron's output as reference. For the binary classification (coherent or thermal), the neuron's output is fed into the identity activation function, and subsequently to the threshold function.
 
%We take as an input-features vector is composed by the first seven probabilities of the photon number distribution, that is, $\bar{P}=\left\{P(0), P(1), P(2), P(3), P(4), P(5), P(6)\right\}$. With this, we achieve that the feature vector size stays fixed for a different number of data points. It is worth mentioning that the determination of an appropriate feature vector is one hard task in machine learning.

To train the ADALINE, we make use of the so-called delta learning rule \cite{Gallant}, in combination with a database of experimentally measured photon-number distributions, considering different mean photon numbers: $\bar{n}=0.44,\;0.53,\;0.67,\;0.77$. The database for each mean photon number was divided into subsets comprising $10, 20,..., 150, 160$ data points. The ADALINE neurons are thus prepared by using one thousand of those subsets, where 70\% are devoted to training and 30\% to testing. In all cases, the training was stopped after 50 epochs.

We have established the baseline performance for our ADALINE neuron by using naive Bayes classifier. This is a simple classifier based on Bayes' theorem \cite{bayesBook}. Throughout this article, we assume that each measurement is independent. Moreover, we represent the measurement of the photon number sequence as a vector $\mathbf{x}=(x_1,...,x_k)$. Then, the probability of this sequence generated from coherent or thermal light is given by $p(C_j|x_1,...,x_k)$, where $C_j$ could denote either coherent or thermal light. Using Bayes' theorem, the conditional probability can be decomposed as $p(C_j|\mathbf{x}) = p(C_j) p(\mathbf{x}| C_j)/p(\mathbf{x})$. By using the chain rule for conditional probability, we have $p(C_k|x_1,...,x_k) = p(C_j) \prod_{i=1}^k p(x_i | C_j)$. Since our light source is either coherent or thermal, we assume $p(C_j)=0.5$. Thus, it is easy to construct a naive Bayes classifier, where one picks the hypothesis with the highest conditional probability $p(C_j|\mathbf{x})$. We used theoretically generated photon-number probability distributions as the prior probability $p(x_i | C_j)$, and used the experimental data as the test data.

\section{Results}

\begin{figure}[t!]
\centering
\includegraphics[width=\linewidth]{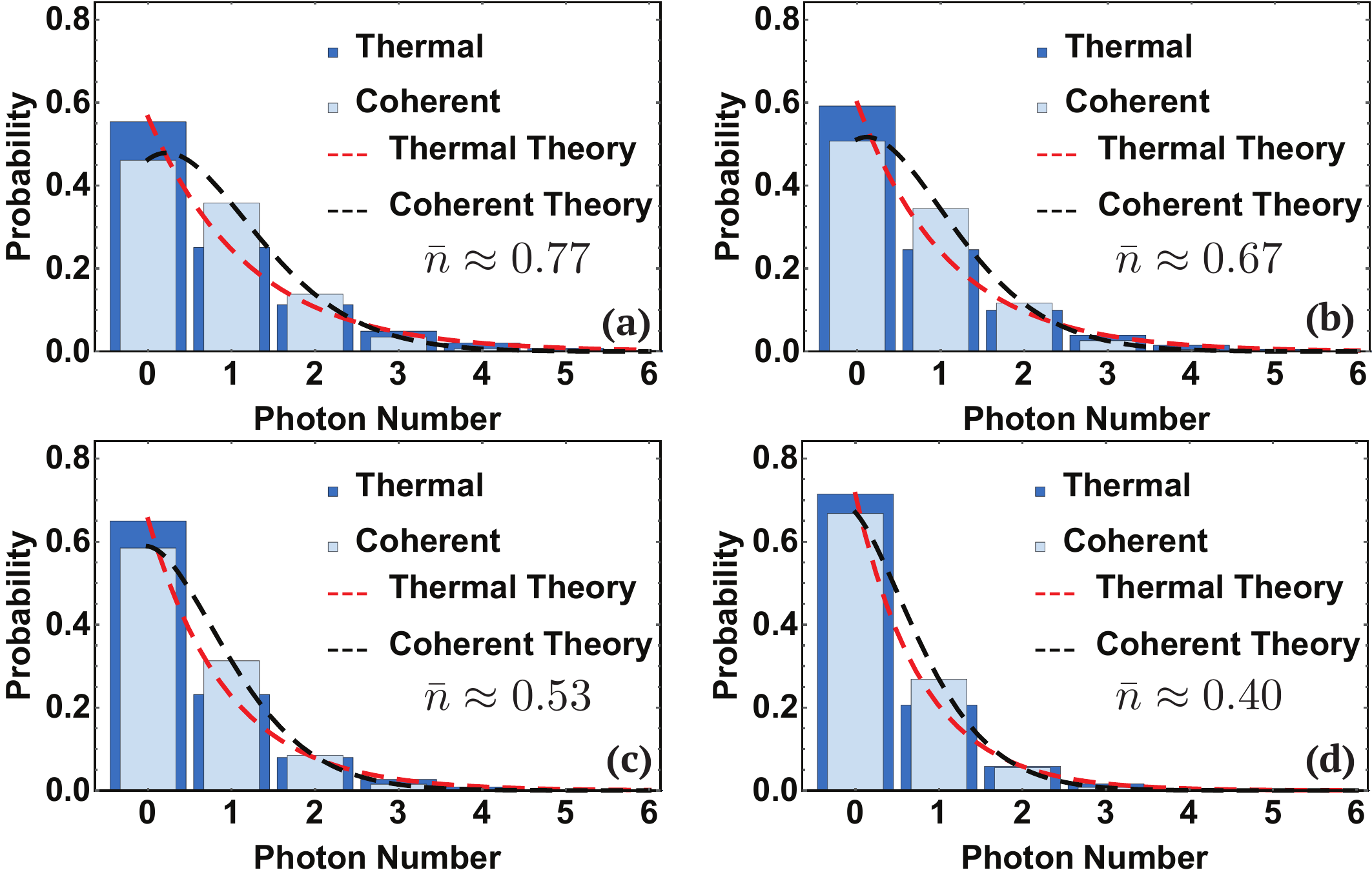}
\caption{A set of histograms displaying theoretical and experimental photon number probability distributions for coherent and thermal light beams with different mean photon numbers. Our experimental results are in excellent agreement with theory. The photon number distributions illustrate the difficulty in discriminating light sources at low-light levels even when large sets of data are available.}
\label{fig:hist_all}
\end{figure}

\begin{figure}[t!]
\centering
\includegraphics[width=0.9\linewidth]{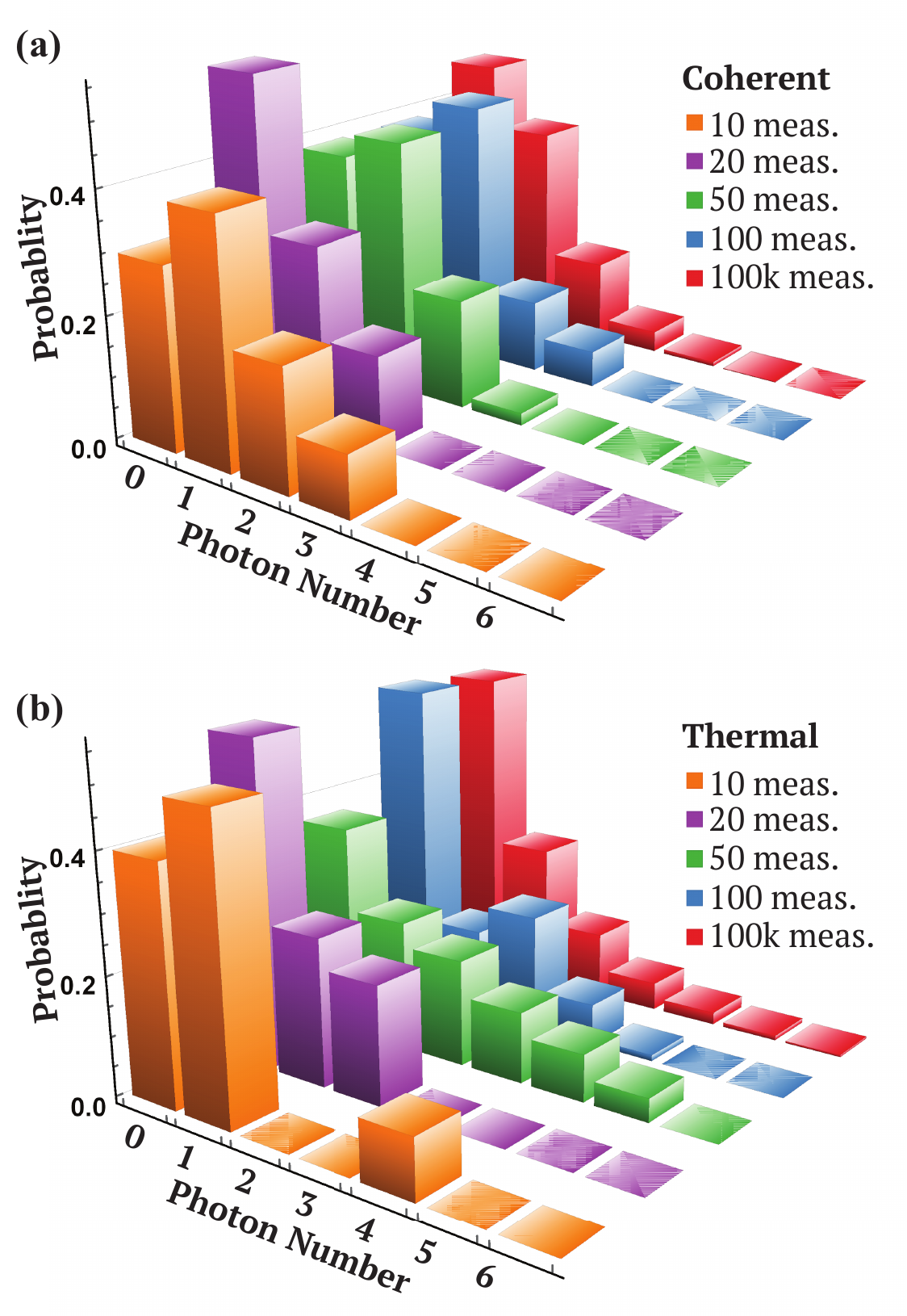}
\caption{Probability distributions of coherent and thermal light, for varying dataset sizes (10, 20, 50, 100, 10k). Data used here is randomly selected from of the measurement presented in Fig. \ref{fig:hist_all} (a).}
\label{fig:hist_num}
\end{figure}

In Fig. \ref{fig:hist_all}, we compare the histograms for the theoretical and experimental photon number distributions for different mean photon numbers $\bar{n}=$ 0.40, 0.53, 0.67 and 0.77. The bar plots are generated by experimental data with one million measurements for each source; the curves in each of the panels represent the expected theoretical photon number distributions for the corresponding mean photon numbers. Fig. \ref{fig:hist_all} shows excellent agreement between theory and experiment which demonstrates the accuracy of our surjective photon counting method. Furthermore, from Fig. \ref{fig:hist_all} (a)-(d), we can also observe the effect of the mean photon number on the photon number probability distributions.
As shown in Fig. \ref{fig:hist_all} (a), it is evident that millions of measurements enable one to discriminate light sources. On the other hand, Fig. \ref{fig:hist_all} (d) shows a situation in which the source mean-photon number is low. In this case, the discrimination of light sources becomes cumbersome, even with millions of measurements. In order to illustrate the difficulty of using limited sets of data to discriminate light sources at low mean photon numbers, we restrict the size of our dataset to 10, 20, 50, 100 and 100000. As shown in Fig. \ref{fig:hist_num}, the photon number distributions obtained with limited number of measurements do not resemble those in
the histograms shown in Fig. \ref{fig:hist_all} (a), for both coherent and thermal light beams.

\begin{figure}[t!]
\centering\includegraphics[width=0.8\linewidth]{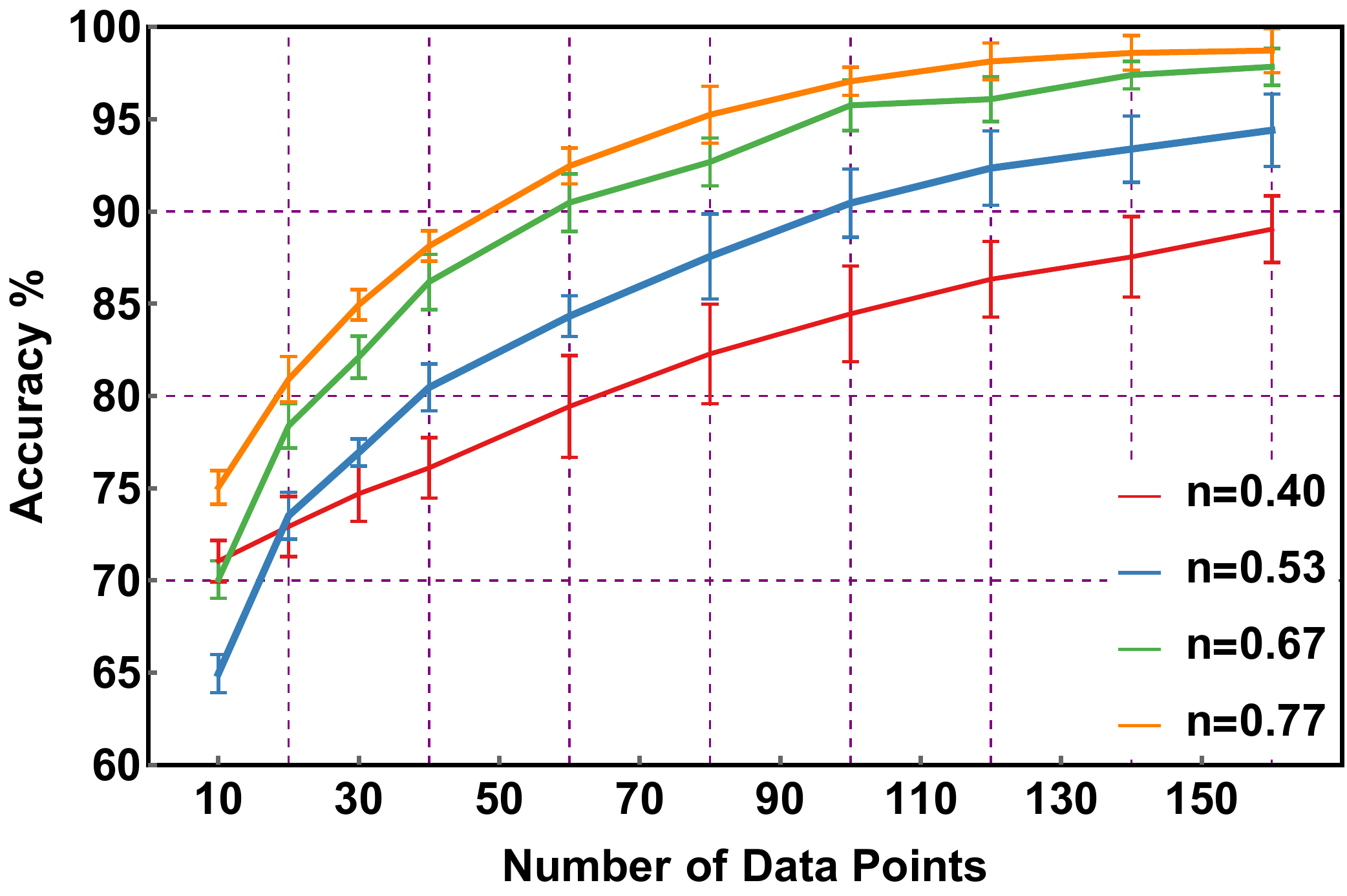}
\caption{Overall accuracy of light discrimination versus the number of data points used in naive Bayes classifier. The curves represent the accuracy of light discrimination for $\bar{n}=0.40$ (red line), $\bar{n}=0.53$ (blue line), $\bar{n}=0.67$ (green line) and $\bar{n}=0.77$ (orange line). The error bars are generated by dividing the data into ten subsets.}
\label{fig:max_accu}
\end{figure}

In Fig. \ref{fig:max_accu}, we show the overall accuracy for light discrimination using naive Bayes classifier. The accuracy increases with the number of data points. For example, when $\bar{n}=0.40$, the accuracy of discrimination increases from approximately 72\% to 88\% as we increase the number of data points from 10 to 160. It is worth noting that even with small increase in number of measurements, the naive Bayes classifier starts to capture the characteristic feature of different light sources, given by distinct sequences of photon number events. This is obvious since larger sets of data contain more information pertaining to the probability distribution. Furthermore, mean photon number of the light field significantly changes the discrimination accuracy profile. As the mean photon number increases, the overall accuracy converges faster towards 100\% as expected. This is due to the fact that the photon number probability distributions become more distinct at higher mean photon number.

\begin{figure}[t!]
\centering
\includegraphics[width=0.8\linewidth]{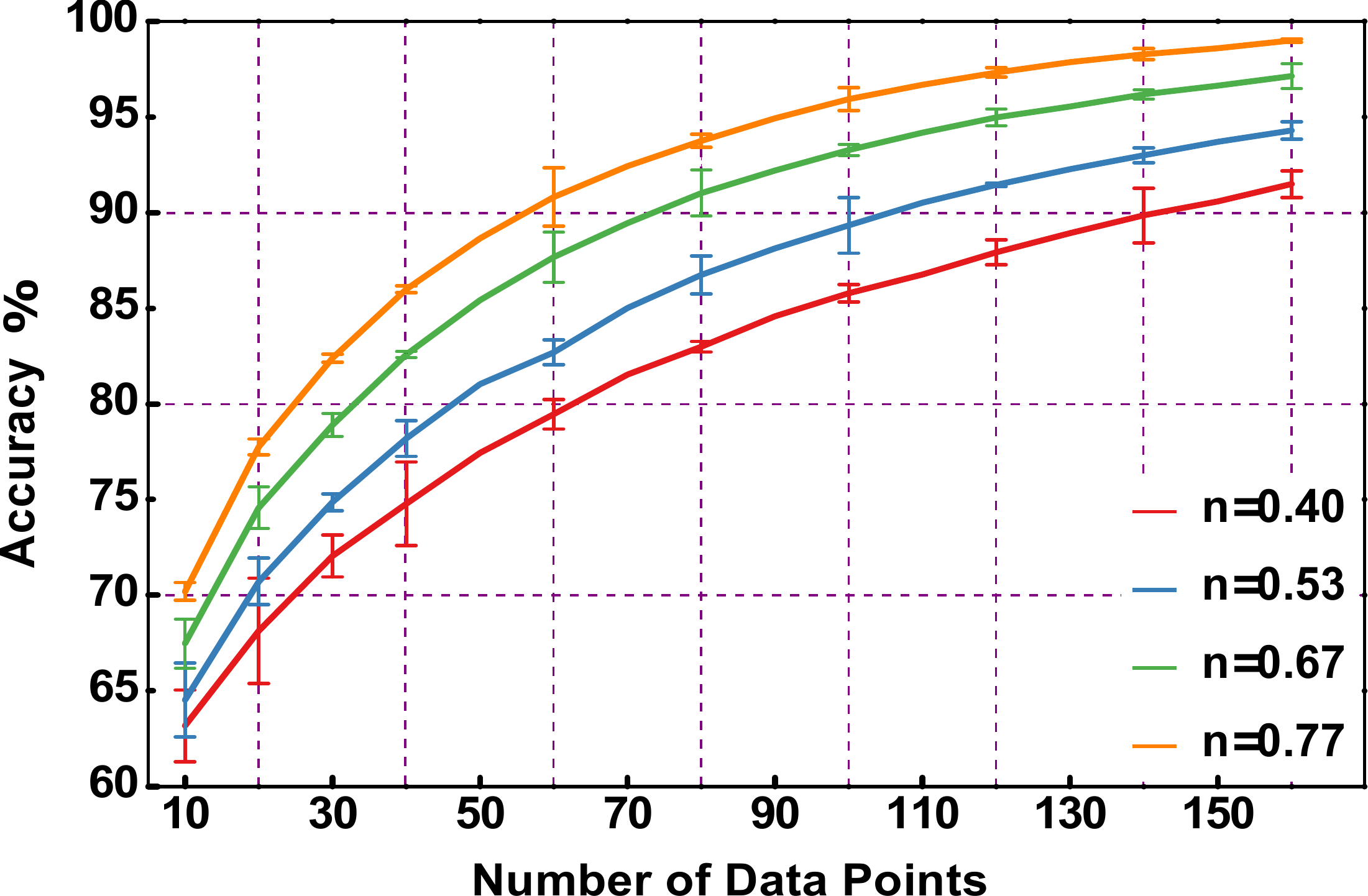}
\caption{Overall accuracy of light discrimination versus the number of data points used in ADALINE. The curves represent the accuracy of light discrimination for $\bar{n}=0.40$ (red line), $\bar{n}=0.53$ (blue line), $\bar{n}=0.67$ (green line) and $\bar{n}=0.77$ (orange line). The error bars represent the standard deviation of the training stages.}
\label{fig:CNN_accu}
\end{figure}

The overall accuracy of light-source discrimination with respect to the number of data points is shown in Fig. \ref{fig:CNN_accu}. Using only 10 data points, ADALINE leads to an average accuracy between 61\%-65\% for $\bar{n}=0.40$; whereas for 160 data points, the accuracy is greater than 90\%. The comparison of Fig. \ref{fig:max_accu} and Fig. \ref{fig:CNN_accu} reveals that ADALINE and naive Bayes classifier exhibit similar accuracy levels. However, ADALINE requires far less computational resources than naive Bayes classifier. As one might expect, in both cases, the accuracy increases with the number of data points and mean photon numbers. Interestingly, the convergence rate for naive Bayes is slightly higher than that of ADALINE classifier. For low mean photon numbers, such as $\bar{n}=0.40$, the improvement in accuracy scales linearly for naive Bayes classifier, as opposed to almost logistic growth that has our ADALINE. This implies that at low mean photon numbers ADALINEs outperform naive Bayes classifier in the sense that the former require much less computational resources than the latter.

\begin{figure}[t!]
\centering
\includegraphics[width=\linewidth]{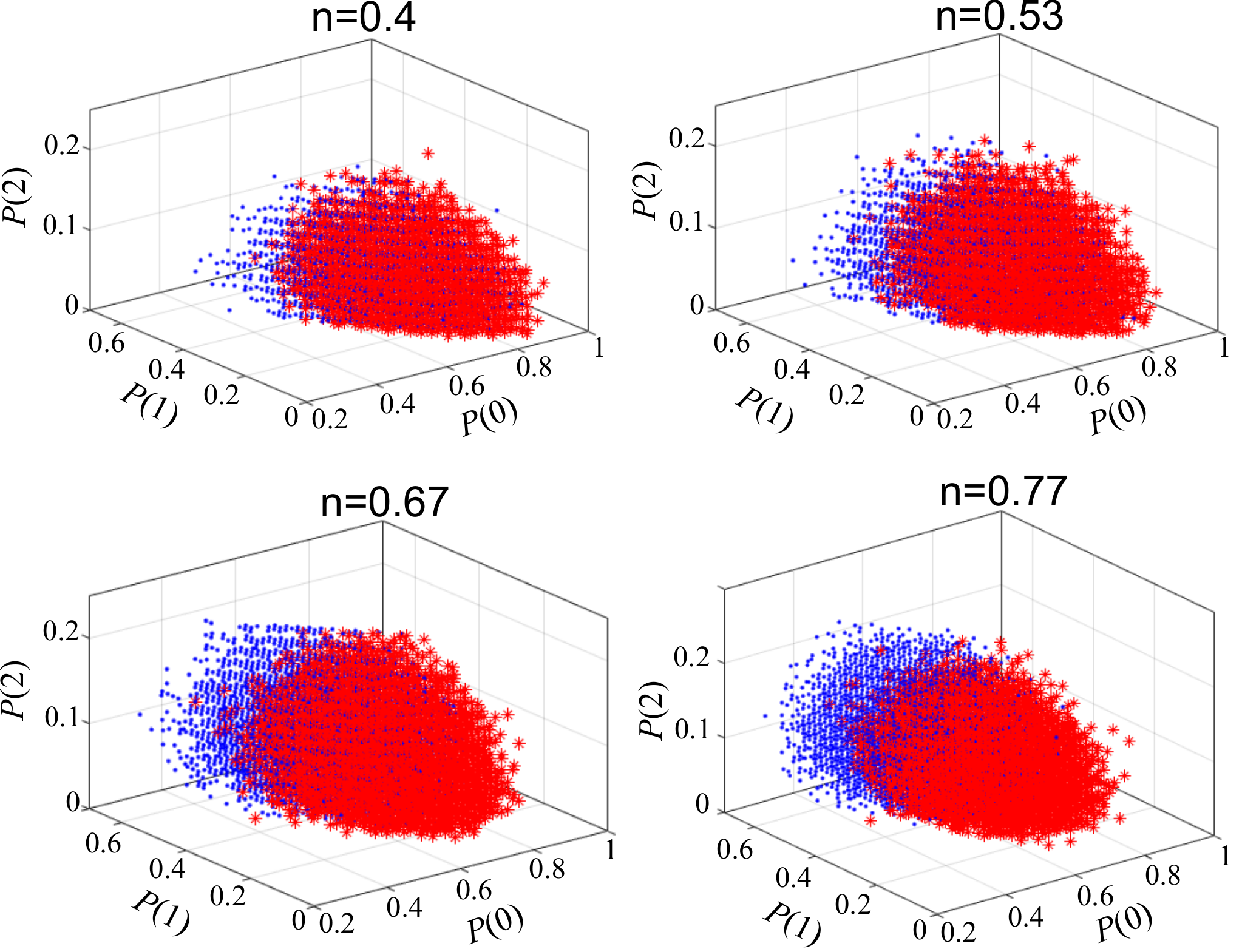}
\caption{Projection of the feature space on the plane ($P(0)$, $P(1)$, $P(2)$) for different mean photon numbers: (a) $\bar{n}=0.4$, (b) $\bar{n}=0.53$, (c) $\bar{n}=0.67$, and (d) $\bar{n}=0.77$. The blue points correspond to photon statistics of coherent light, whereas the red stars describe photon statistics of thermal light. In all cases the number of data points is fixed at $M=60$.}
\label{fig:numberphoton}
\end{figure}

\begin{figure}[t!]
\centering
\includegraphics[width=\linewidth]{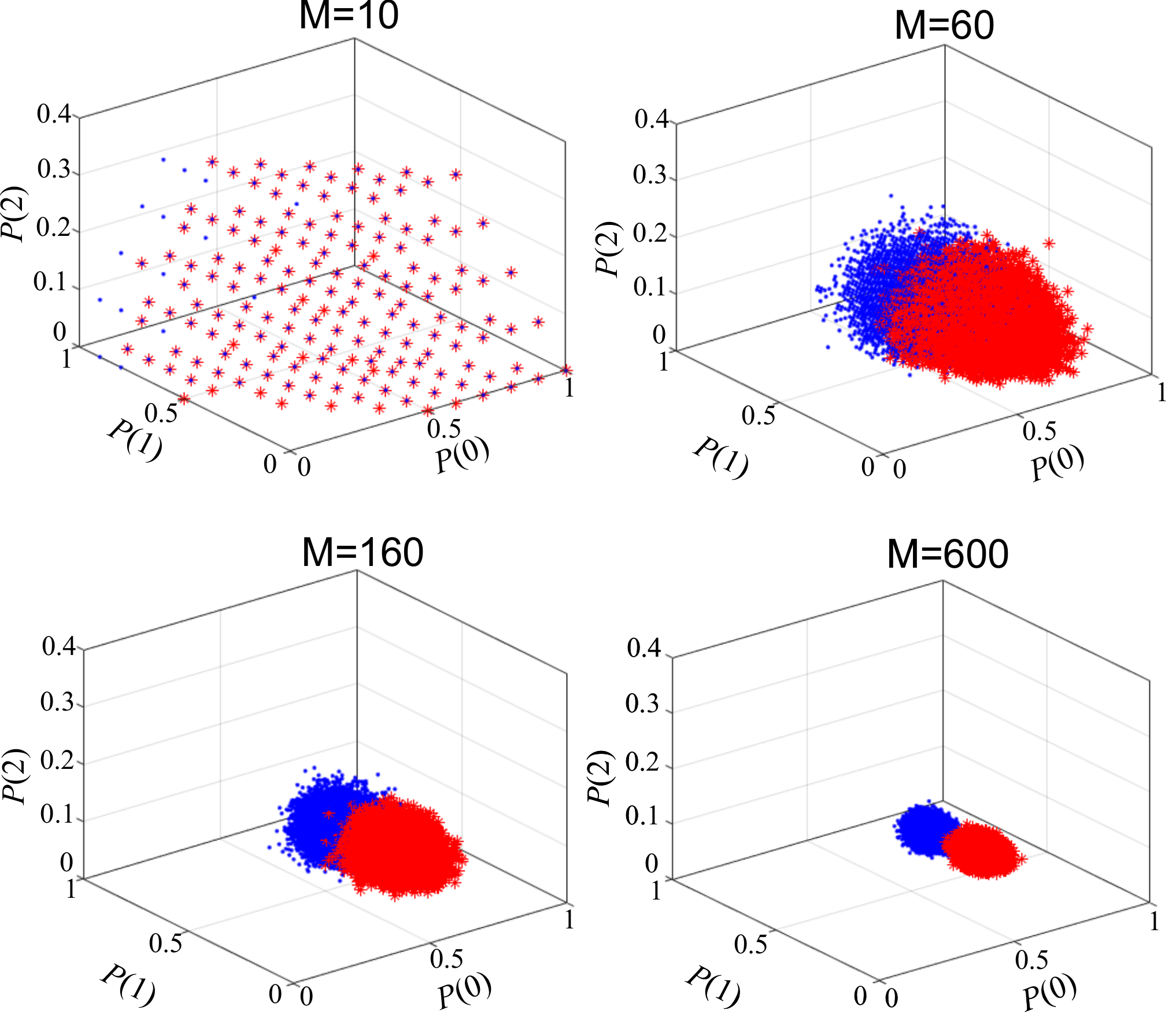}
\caption{Projection of the feature space on the plane ($P(0)$, $P(1)$, $P(2)$) for different number of data points: (a) $M=10$, (b) $M=60$, (c) $M=160$, and (d) $M=600$. The blue points correspond to photon statistics of coherent light, whereas the red stars describe photon statistics of thermal light. In all cases, the mean photon number is set to $\bar{n}=0.77$}
\label{fig:numberdatapoint}
\end{figure}

To understand why a single ADALINE neuron is enough for light discrimination, we first realize that ADALINE is a linear classifier; therefore the decision surface is expressed by a seven-dimensional hyper-plane, defined by the seven $P(n)$ (with $n=0,1,...,6$) features. Interestingly, one can find that the datasets at the space of probability-distribution values are linearly separable. This can be seen from Fig. \ref{fig:numberphoton}, where we plot the projection of the feature space on a three-dimensional sub-space defined by ($P(0)$, $P(1)$, $P(2)$) considering different mean photon numbers $\bar{n}=0.4,0.53,0.67$ and $0.77$. In all cases, the number of data points is fixed at $M=60$. Within this subspace, the datasets corresponding to the photon statistics of thermal (red stars) and coherent (blue points) lights separate each other as $\bar{n}$ increases. This effect is more evident when the number of data points is increased, and the mean photon number remains fixed at $\bar{n}=0.77$ [see Fig. \ref{fig:numberdatapoint}]. Evidently, the fact that both, thermal and coherent light form two well linearly separated classes makes ADALINE the optimum classifier for light identification.

%Note that for an excessive number of data points $M=600$, the datasets are clearly separated. As $M$ increases the feature set tends to focus on a particular region. By contrast, when $M$ is low, the datasets disperse, making one hard task to discriminate them. Nevertheless, we have shown that an ADALINE and naive Bayes classifiers are capable of discriminating light sources at low-light levels even when large sets of data are available.

%This implies that the datasets at the space of probability distributions are linearly separable, because the classes can be correctly discerned by ADALINE.

\section{Conclusion}
For more than twenty years, there has been an enormous interest in reducing the number of photons and measurements required to perform imaging, remote sensing and metrology at extremely low-light levels \cite{dowling:93, maganab:19}. In this regard, photonic technologies operating at low-photon levels utilize weak photon signals that make them vulnerable against detection of  environmental photons emitted from natural sources of light. Indeed, this limitation has made unfeasible the realistic implementation of this family of technologies \cite{sher:18, hlousek:19, howard:19}. So far, this vulnerability has been tackled through conventional approaches that rely on the measurement of coherence functions, the implementation of thresholding and quantum state tomography \cite{sher:18, hlousek:19, howard:19, cohen:19}. Unfortunately, these approaches to characterize photon-fluctuations rely on the acquisition of large number of measurements that impose constraints on the identification of light sources. Here, for the first time, we have demonstrated a smart protocol for discrimination of light sources at mean photon numbers below one. Our work demonstrates a dramatic improvement of several orders of magnitude in both the number of photons and measurements  required to identify light sources \cite{sher:18, hlousek:19, howard:19, cohen:19}. Furthermore, our results indicate that a single artificial neuron outperforms naive Bayes classifier at low-light levels. Interestingly, this neuron has simple analytical and computational properties that enable low-complexity and low-cost implementations of our technique. We are certain that our work has important implications for multiple photonic technologies, such as LIDAR and microscopy of biological materials.

\section*{Acknowledgements}
We thank the Department of Physics \& Astronomy at Louisiana State University for providing Startup funding to perform this experimental work. CY would like to acknowledge support from National Science Foundation. NB would like to thank Army Research Office (ARO) for the funding. R.J.L.M. and M.A.Q.J. thankfully acknowledge financial support by CONACYT under the project CB-2016-01/284372 and by DGAPA-UNAM under the project UNAM-PAPIIT IA100718. We all thank J. P. Dowling, X. Wang, L. Cohen and H. S. Eisenberg for helpful discussions.

%Manual citation list

\newpage

\begin{widetext}
\begin{flushleft}
\singlespacing{\Large{\textbf{Supplementary material:\vspace{2mm}\\
\Large{Identification of Light Sources using using Machine Learning}}}}
\end{flushleft}

\vspace{0.5cm}

\noindent In this supplemental material, we evaluate two additional machine-learning algorithms, namely a one-dimensional convolutional neural network (1D CNN) and a multilayer neural network (MNN). Despite both algorithms are effective to identify light sources, they are analytically and computationally more sophisticated than the simple ADALINE model, but their recognition rates do not present substantial differences. Figures \ref{fig:models}(a) and  \ref{fig:models}(b) show the structure of the 1D-CNN and MNN, respectively.

\begin{figure}[htbp]
\centering
\includegraphics[width=16cm]{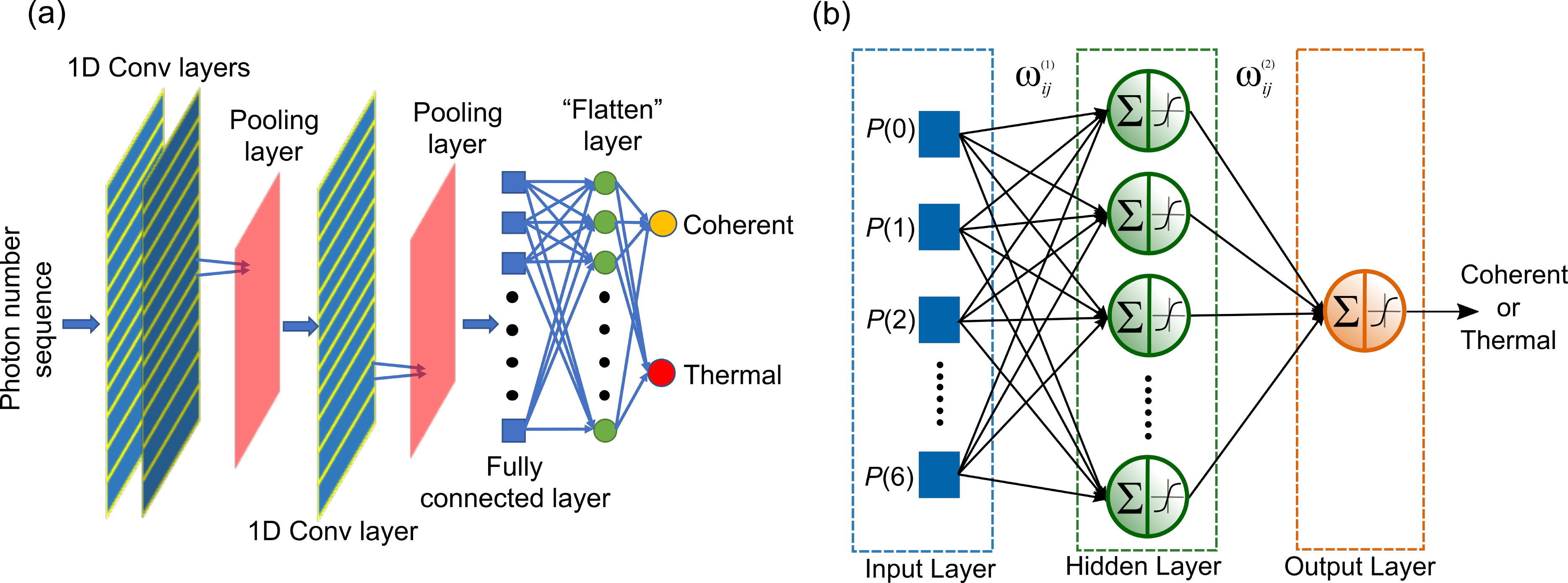}
\caption{Schematic representations: (a) one-dimensional convolutional neural network and (b) multilayer neural network used for demonstration of light source identication.}
\label{fig:models}
\end{figure}

A convolutional neural network is a deep learning algorithm that extracts automatically relevant features of the input \cite{Ian}. Here, our one-dimensional convolutional neural network is composed by two 1D-convolutional layers that extract the low and high-level features of the input. Outcomes from these two layers are subsequently fed into a convolutional layer sandwiched between two max-pooling layers. The pooling layers downsample the input representation, and therefore its dimensionality, leading to a computational simplification by removing redundant and unnecessary information. The activation function, implemented in all layers, is the rectified linear unit function (ReLU). Finally, a fully connected and a flattening layer precedes the output layer consisting of two softmax functions, whose outputs are the probability distributions over labels.

\begin{figure}[htbp]
\centering
\includegraphics[width=17cm]{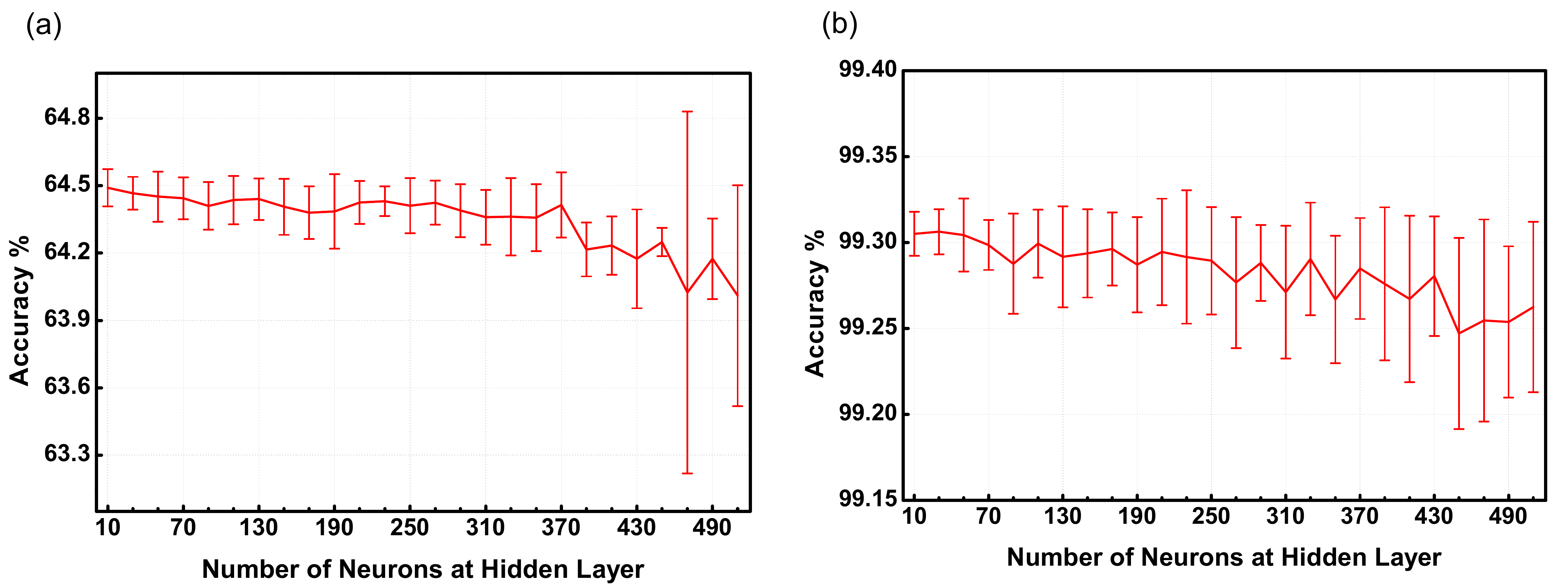}
\caption{Overall accuracy of light discrimination versus the number of neurons in the hidden layer of the MNN by considering two different mean photon numbers: (a) $\bar{n}=0.4$ and (b) $\bar{n}=0.77$. The error bars represent the standard deviation of the training stages.}
\label{fig:number}
\end{figure}

On the other hand, the multilayer neural network belongs to a classical machine learning algorithm, where the feature vector should be manually determined \cite{Bishop}. In our case, this vector is given by the probabilities of the photon number distribution, $P(n)$. As depicted in Fig. \ref{fig:models}(b), the model corresponds to a two-layer feed-forward network: the hidden layer contains ten sigmoid neurons and the output layer consists of a softmax function.
To determine a suitable neuron number in the hidden layer of the MNN, we trained different MNNs by changing the neuron number in the hidden layer and followed the accuracy values for each net. Figs.  \ref{fig:number}(a) and  \ref{fig:number}(b) show the overall accuracy for light discrimination versus the number of neurons in the hidden layer for different mean photon numbers, $\bar{n}=0.4$ and $0.77$, respectively. Note that in both cases, the accuracy becomes lower as the number of neurons increases. This is because many neurons lead to over-parameterization, causing poor generalization of the test-stage data. Additionally, as the number of neurons increases, the training becomes computationally more intensive. All the MNNs were trained by using the scaled conjugate gradient backpropagation method where the cross-entropy was employed as the cost function. Since the output of sigmoid neurons is ranged in the interval [0,1], the cross-entropy function is ideal for the classification task. The network training was stopped after 200 epochs.

1D-CNNs and MNNs were trained using the same training set described in the main manuscript. Despite that deep neural network should be trained with a larger amount of data, we use 70\% of the dataset for the training and the rest for testing both networks. Note that the same procedure was used for the ADALINE model. Figures \ref{fig:accuracy}(a) and \ref{fig:accuracy}(b) show the overall light-discrimination accuracy for increasingly larger number of data points for (a) 1D-CNNs and (b) MNNs. In both cases, the accuracy increases with the number of data points, because larger sets of data contain more information about the probability distribution. Interestingly, the accuracy of 1D-CNNs for $\bar{n}=0.67$ and $\bar{n}=0.77$ are almost the same; this indicates that in the low mean photon-number regime, the peak performance for 1D-CNN saturates much faster than the MNN classifier. As one might expect, this fast accuracy convergence carries the cost of a much complex computation as compared to the one needed for the MNN classifier.

\begin{figure}[htbp]
\centering
\includegraphics[width=16cm]{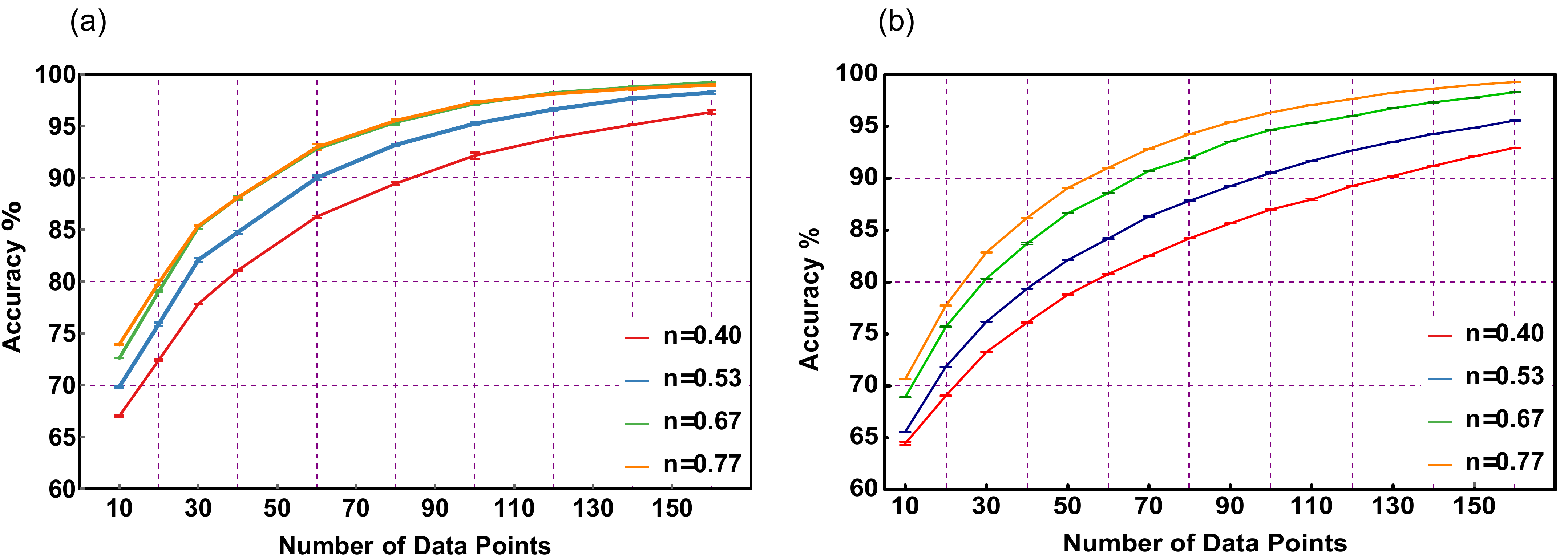}
\caption{Overall accuracy of light discrimination versus the number of data points used in (a) 1D-CNN and (b) MNN. The curves represent the accuracy of light discrimination for $\bar{n}=0.40$ (red line), $\bar{n}=0.53$ (blue line), $\bar{n}=0.67$ (green line) and $\bar{n}=0.77$ (orage line). The error bars represent the standard deviation of the training epochs for 1D-CNN and training stages for MNN.}
\label{fig:accuracy}
\end{figure}

\section*{References}
\vspace{-0.5cm}
\renewcommand{\refname}{\large{}}

\end{widetext}


\begin{thebibliography}{1}
\newcommand{\enquote}[1]{``#1''}

\bibitem{glauber:63}
R. J. Glauber,
\enquote{The Quantum Theory of Optical Coherence,} Phys. Review \textbf{130},
2598 (1963).

 \bibitem{mandel:95}
L. Mandel and E. Wolf
\enquote{Optical Coherence and Quantum optics,} Cambridge: Cambridge University Press (1995).

 \bibitem{mandel:79}
L. Mandel,
\enquote{Sub-Poissonian photon statistics in resonance fluorescence,} Opt. Lett. \textbf{4},
205--207 (1979).

 \bibitem{mandel:65}
L. Mandel and E. Wolf,
\enquote{Coherence properties of optical fields,} Rev. Mod. Phys. \textbf{37},
231 (1965).

 \bibitem{liu:09}
J. Liu, and Y. Shih,
\enquote{Nth-order coherence of thermal light,} Phys. Rev. A \textbf{79},
023819 (2009).

 \bibitem{hlousek:19}
J. Hlousek, M. Dudka, I. Straka, and M. Jazek,
\enquote{Accurate Detection of Arbitrary Photon Statistics,} Phys. Rev. Lett. \textbf{123},
153604 (2019).

\bibitem{dovrat:12}
L. Dovrat, M. Bakstein, D. Istrati, A. Shaham, and H. S. Eisenberg, \enquote{Measurements of the dependence of the photon-number distribution on the number of modes in parametric down-conversion,} Opt. Express 20, 2266-2276 (2012)

 \bibitem{dovrat:13}
L. Dovrat, M. Bakstein, D. Istrati, E. Megidish, A. Halevy, L. Cohen and H. S. Eisenberg,
\enquote{Direct observation of the degree of correlations using photon-number-resolving detectors,} Phys. Rev. A \textbf{87},
053813  (2013).

 \bibitem{zambra:05}
G. Zambra, A. Andreoni, M. Bondani, M. Gramegna, M. Genovese, G. Brida, A. Rossi, and M. G. A. Paris,
\enquote{Experimental Reconstruction of Photon Statistics without Photon Counting,} Phys. Rev. Lett. \textbf{95},
063602   (2005).

 \bibitem{howard:19}
L. A. Howard, G. G. Gillett, M. E. Pearce, R. A. Abrahao, T. J. Weinhold, P. Kok, and A. G. White,
\enquote{Optical Imaging of Remote Bodie using Quantum Detectors,} Phys. Rev. Lett. \textbf{123},
143604  (2019).

%  \bibitem{Donovan:93}
% D. P. Donovan, J. A. Whiteway, and A.I. Carswell,
% \enquote{Correction for nonlinear photon-counting effects in lidar systems,} App. Opt. \textbf{32},
% 6742-6753 (1993).

 \bibitem{dowling:93}
J. P. Dowling, and K. P. Seshadreesan,
\enquote{Quantum Optical Technologies for Metrology, Sensing, and Imaging,} J. Light. Technol. \textbf{33},
2359 (2015).

 \bibitem{sher:18}
Y. Sher, L. Cohen, D. Istrati, and H. S. Eisenberg,
\enquote{Low intensity lidar using compressed sensing and a photon number resolving detector,} Emerging Digital Micromirror Device Based Systems and Applications X \textbf{10546},
105460J (2018).

 \bibitem{wang:16}
Q. Wang, L. Hao, Y. Zhang, C. Yang, X. Yang, L. Xu, and Y. Zhao,
\enquote{Optimal detection strategy for super-resolving quantum lidar,} J. Appl. Phys. \textbf{119},
023109 (2016).

 \bibitem{dowling:08}
J. P. Dowling,
\enquote{Quantum optical metrology -- the lowdown on High-N00N states,} Contemp. Phys \textbf{49},
125-143  (2008).

 \bibitem{Omar:19}
O. S. Magana-Loaiza, R. J. Leon-Montiel, A. Perez-Leija, A. B. URen, C. You, K. Busch, A. E. Lita, S. W. Nam, R. P. Mirin, T. Gerrits,
\enquote{Multiphoton Quantum-State Engineering using Conditional Measurements,} npj Quantum Information \textbf{5}, 80 (2019).

 \bibitem{lecun:15}
Y. LeCun, Y. Bengio, and G. Hinton,
\enquote{Deep learning,} Nature \textbf{521},
436-444 (2015).

 \bibitem{biamonte:17}
J. Biamonte, P. Wittek, N. Pancotti, P. Rebentrost, N. Wiebe, and S. Lloyd,
\enquote{Quantum Machine Learning,} Nature \textbf{549},
195-202 (2017).

 \bibitem{Dunjko:16}
V. Dunjko, J. M. Taylor, H. J. Briegel,
\enquote{Quantum-Enhanced Machine Learning,} Phys. Rev. Lett.  \textbf{117},
130501 (2016).

 \bibitem{Hentschel:10}
A. Hentschel, and B. C. Sanders,
\enquote{Machine Learning for Precise Quantum Measurement,} Phys. Rev. Lett.  \textbf{104},
063603 (2010).

 \bibitem{Lumino:18}
A. Lumino, E. Polino, A. S. Rab, G. Milani, N. Spagnolo, N. Wiebe, and F. Sciarrino,
\enquote{Experimental Phase Estimation Enhanced by Machine Learning,} Phys. Rev. Appl.  \textbf{10},
044033 (2018).

\bibitem{Zeilinger:18}
A. A. Melnikov, H. P. Nautrup, M. Krenn, V. Dunjko, M. Tiersch, A. Zeilinger and H. J. Briegel, \enquote{Active learning machine learns to create new quantum experiments,} PNAS \textbf{6}, 1221-1226 (2018).

 \bibitem{Lohani:18}
S. Lohani, R. T. Glasser,
\enquote{Turbulence correction with artificial neural networks,} Opt. Lett.  \textbf{43},
2511-2614 (2018).

 \bibitem{Gao:18}
J. Gao, L.-F. Qiao, Z.-Q. Jiao, Y.-C. Ma, C.-Q. Hu, R.-J. Ren, A.-L. Yang, H. Tang, M.-H. Yung, and X.-M. Jin,
\enquote{Experimental Machine Learning of Quantum States,} Phys. Rev. Lett.  \textbf{120},
240501 (2018).

 \bibitem{Torlai:18}
G. Torlai, G. Mazzola, J. Carrasquilla, M. Troyer, R. Melko, and G. Carleo,
\enquote{Neural-network quantum state tomography,} Nat. Phys.  \textbf{14},
447-450 (2018).

\bibitem{surjective}
S. M. H. Rafsanjani, M. Mirhosseini, O. S. Maga\~na-Loaiza, B. T. Gard, R. Birrittella, B. E. Koltenbah, C. G. Parazzoli, B. A. Capron, C. C. Gerry, J. P. Dowling, and R. W. Boyd, \enquote{Quantum-enhanced interferometry with weak thermal light,} Optica \textbf{4}, 487-491 (2017).

\bibitem{Burenkov:17}
I. A. Burenkov, A. K. Sharma, T. Gerrits, G. Harder, T. J. Bartley, C. Silberhorn, E. A. Goldschmidt, and S. V. Polyakov, \enquote{Full statistical mode reconstruction of a light field via a photon-number-resolved measurement}, Phys. Rev. A. \textbf{95}, 053806 (2017)

\bibitem{Montaut:18}
N. Montaut, O. S. Magaña-Loaiza, T. J. Bartley, V. B. Verma, S. W. Nam, R. P. Mirin, C. Silberhorn, and T. Gerrits, \enquote{Compressive characterization of telecom photon pairs in the spatial and spectral degrees of freedom}, Optica \textbf{11}, 1418 (2018)

\bibitem{Bernard}
B. Windrow, M. E. Hoff, \enquote{Adaptive switching circuits,} Technical Report No. 1553-1, Stanford University, Stanford-California, Stanford Electronics Laboratories (1960).

\bibitem{Gallant}
S. I. Gallant, \enquote{Neural network learning and expert systems,} MIT press (1993).

\bibitem{Cauchy}
A. Cauchy, \enquote{M\'ethode g\'en\'erale pour la r\'esolution des systemes d’\'equations simultan\'ees,} Comp. Rend. Sci. Paris 25.1847, pp. 536-538 (1847).

\bibitem{bayesBook}
A. Gelman, J. B. Carlin, H. S. Stern, D. B. Dunson, A. Vehtari, D. B. Rubin, \enquote{Bayesian Data Analysis,} Chapman and Hall/CRC (2013).

\bibitem{maganab:19}
O. S. Magana-Loaiza, and R. W. Boyd, \enquote{Quantum Imaging and Information,} Reports on Progress in Physics, in press (2019).

\bibitem{cohen:19}
L. Cohen, E. S. Matekole, Y. Sher, D. Istrati, H. S. Eisenberg, and J. P Dowling,  \enquote{Thresholded Quantum LIDAR - Exploiting Photon-Number-Resolving Detection}, Phys. Rev. Lett., in press (2019)

\end{thebibliography}

\begin{thebibliography}{XX}

\bibitem{Ian} Goodfellow, Ian, Yoshua Bengio, and Aaron Courville. Deep learning. MIT press, 2016.

\bibitem{Bishop} Bishop, Christopher M. Pattern recognition and machine learning. Springer, 2006.

\end{thebibliography}
\end{document}